\documentclass[aps, 
	prapplied, 
	noeprint, 
	superscriptaddress, 
	floatfix, 
	nofootinbib, 
	tightenlines, 
	twocolumn
	]{revtex4-2}

\usepackage{graphicx} 
\usepackage{dcolumn} 
\usepackage{bm}
\usepackage{color}
\usepackage{float}
\usepackage{dsfont}
\usepackage{subfig}
\usepackage[breaklinks=true]{hyperref}
\usepackage{listings}
\usepackage{braket}
\usepackage{siunitx}  
\usepackage{ragged2e}  
\usepackage[english]{babel}
\hypersetup{
    bookmarks=true,         
    unicode=false,         
    pdftoolbar=true,         
    pdfmenubar=true,       
    pdffitwindow=false,     
    pdfstartview={FitH},    
    pdfauthor={Soeren Wengerowsky},    
    colorlinks=true,    
    linkcolor=blue,         
    citecolor=red,        
}

\makeatletter
\renewcommand{\fnum@figure}{\textbf{FIG. \thefigure}}
\renewcommand{\fnum@table}{\textbf{TABLE \thetable}}
\renewcommand{\@makecaption}[2]{%
  \vskip\abovecaptionskip
  \justifying\normalsize\rmfamily #1. #2\par
  \vskip\belowcaptionskip}
\makeatother

\graphicspath{{figures/}}

\newcommand{\PrYSO}{Pr$^{3+}$:Y$_2$Si{O$_5\,$}}

\begin{document}

\definecolor{dkgreen}{rgb}{0,0.6,0}
\definecolor{gray}{rgb}{0.5,0.5,0.5}
\definecolor{mauve}{rgb}{0.58,0,0.82}

\lstset{frame=tb,
  	language=Matlab,
  	aboveskip=3mm,
  	belowskip=3mm,
  	showstringspaces=false,
  	columns=flexible,
  	basicstyle={\small\ttfamily},
  	numbers=none,
  	numberstyle=\tiny\color{gray},
 	keywordstyle=\color{blue},
	commentstyle=\color{dkgreen},
  	stringstyle=\color{mauve},
  	breaklines=true,
  	breakatwhitespace=true
  	tabsize=3
}

\title{Quantum Frequency Conversion of $\mu s$-long Photons from the Visible to the Telecom-C-Band}
\author{S\"{o}ren Wengerowsky$^{*{\dagger}}$}
\affiliation{ICFO-Institut de Ciencies Fotoniques, The Barcelona Institute of Technology, Mediterranean Technology Park, 08860 Castelldefels (Barcelona), Spain}
\author{Stefano Duranti$^*$}
\affiliation{ICFO-Institut de Ciencies Fotoniques, The Barcelona Institute of Technology, Mediterranean Technology Park, 08860 Castelldefels (Barcelona), Spain}
\author{Lukas Heller}
\affiliation{ICFO-Institut de Ciencies Fotoniques, The Barcelona Institute of Technology, Mediterranean Technology Park, 08860 Castelldefels (Barcelona), Spain}

\author{Hugues de Riedmatten$^{\ddagger}$}
\affiliation{ICFO-Institut de Ciencies Fotoniques, The Barcelona Institute of Technology, Mediterranean Technology Park, 08860 Castelldefels (Barcelona), Spain}
\affiliation{ICREA-Instituci\'{o} Catalana de Recerca i Estudis Avan\c cats, 08015 Barcelona, Spain}
\date{\today}

\begin{abstract}
Quantum Frequency Conversion (QFC) is a widely used technique to interface atomic systems with the telecom band in order to facilitate propagation over longer distances in fiber. Here we demonstrate the difference-frequency conversion from \qty{606}{\nm} to \qty{1552}{\nm} of microsecond-long weak coherent pulses at the single photon level compatible with \PrYSO quantum memories, with high-signal to noise ratio. We use a single step difference frequency generation process with a continuous-wave pump at \qty{994}{\nm} in a periodically poled Lithium Niobate (MgO:ppLN) waveguide and  ultra-narrow spectral filtering down to a bandwidth of \qty{12.5}{\MHz}.  With this setup, we achieve the conversion of weak coherent pulses of duration up to \qty{13.6}{\micro\second} with a device efficiency of about \qty{25}{\percent} and a signal-to-noise ratio $>$\num{460} for  \qty{10}{\micro\second}-long pulses containing one photon on average.  This signal-to-noise ratio is large enough to enable a high-fidelity conversion of qubits emitted from an emissive quantum memory based on \PrYSO and to realize an interface with quantum processing nodes based on narrow-linewidth cavity-enhanced trapped ions. 
\end{abstract}

\maketitle

\def\thefootnote{*}\footnotetext{These authors contributed equally to this work.}\def\thefootnote{\arabic{footnote}}
\def\thefootnote{$\dagger$}\footnotetext{\href{mailto:soeren.wengerowsky@icfo.eu}{soeren.wengerowsky@icfo.eu}}\def\thefootnote{\arabic{footnote}}
\def\thefootnote{$\ddagger$}\footnotetext{\href{mailto:hugues.deriedmatten@icfo.eu}{hugues.deriedmatten@icfo.eu}}\def\thefootnote{\arabic{footnote}}

\section{Introduction}
Quantum communication is a rapidly developing field that has the potential to revolutionize secure communication and information processing. An important challenge is to distribute entanglement between remote quantum nodes to form a quantum network~\cite{kimble2008quantum,Wehner2018}. Such a network may include different kinds of quantum nodes with different functionalities, consisting of  very different physical systems operating  at different wavelengths, often in the visible range. One example of such a hybrid architecture is the distribution of entanglement between remote quantum processing nodes using a quantum repeater chain based on multiplexed quantum memories. In that case, the entanglement would first be distributed over long distance using the quantum repeater chain, and then transferred to the processing nodes. This last step requires that the two different types of nodes emit indistinguishable single photons such that they can interfere, which involves the use of a quantum frequency conversion step~\cite{kumar1990quantum}. 

Several physical systems have been investigated as quantum network nodes~\cite{Lei2023}, including cold atomic gases and rare-earth doped solids for multiplexed quantum memories and individual trapped ions, Rydberg atoms or color centers in diamond for quantum processing nodes. Trapped ions in cavities (e.g. based on calcium ions) have demonstrated outstanding performance as quantum processing nodes~\cite{Postler2022,hucul2015,krutyanskiy2023,Krutyanskiy2023a,krutyanskiy2024}. However, when embedded in high-finesse cavities to enhance the interaction with light, the photons generated have usually a narrow linewidth of the order of 100 kHz, leading to very long photon duration of several microseconds~\cite{Stute2012,Krutyanskiy2019}. To include these kind of nodes in a hybrid network, it is therefore necessary to develop a quantum frequency converter that allows conversion of such long photons with high fidelity, either with direct conversion from one node to the other, or from each node to a common wavelength allowing quantum state transfer via quantum teleportation using a Bell state measurement. For the latter scenario, a convenient common wavelength is the telecom C band, where the loss in optical fibers is minimal, allowing long distance operation.   

Several important quantum nodes systems emit light around \qty{600}{\nm}, e.g. europium doped quantum memories (\qty{580}{\nm}), praseodymium doped quantum memories (\qty{606}{\nm}), nitrogen (\qty{632}{\nm}) and tin (\qty{619}{\nm}) vacancy centers in diamond. Praesodymium-doped crystals such as \PrYSO, are a promising candidate for multiplexed~\cite{lago-rivera2021telecomheralded,Yang2018b, seri2019quantum, ortu2022multimode} and efficient~\cite{hedges2010efficient,duranti2024c} quantum memories due to their long coherence times \cite{heinze2013stopped}, high optical depth and large inhomogeneous broadening. The systems can be used to store external single photons, using e.g. the atomic frequency comb protocol~\cite{afzelius2009multimode,rakonjac2021entanglement}, but also  provide a good platform for emissive quantum memories, emitting $\mu s$-long single photons at \qty{606}{\nm} entangled with spin waves~\cite{ferguson2016generation,kutluer2017solidstate,laplane2017multimode,kutluer2019time}. In this case as well, frequency conversion to telecom wavelength is needed to transmit the photons over long distances in optical fiber. 
\begin{figure*}[ht!]
\includegraphics[width=2\columnwidth]{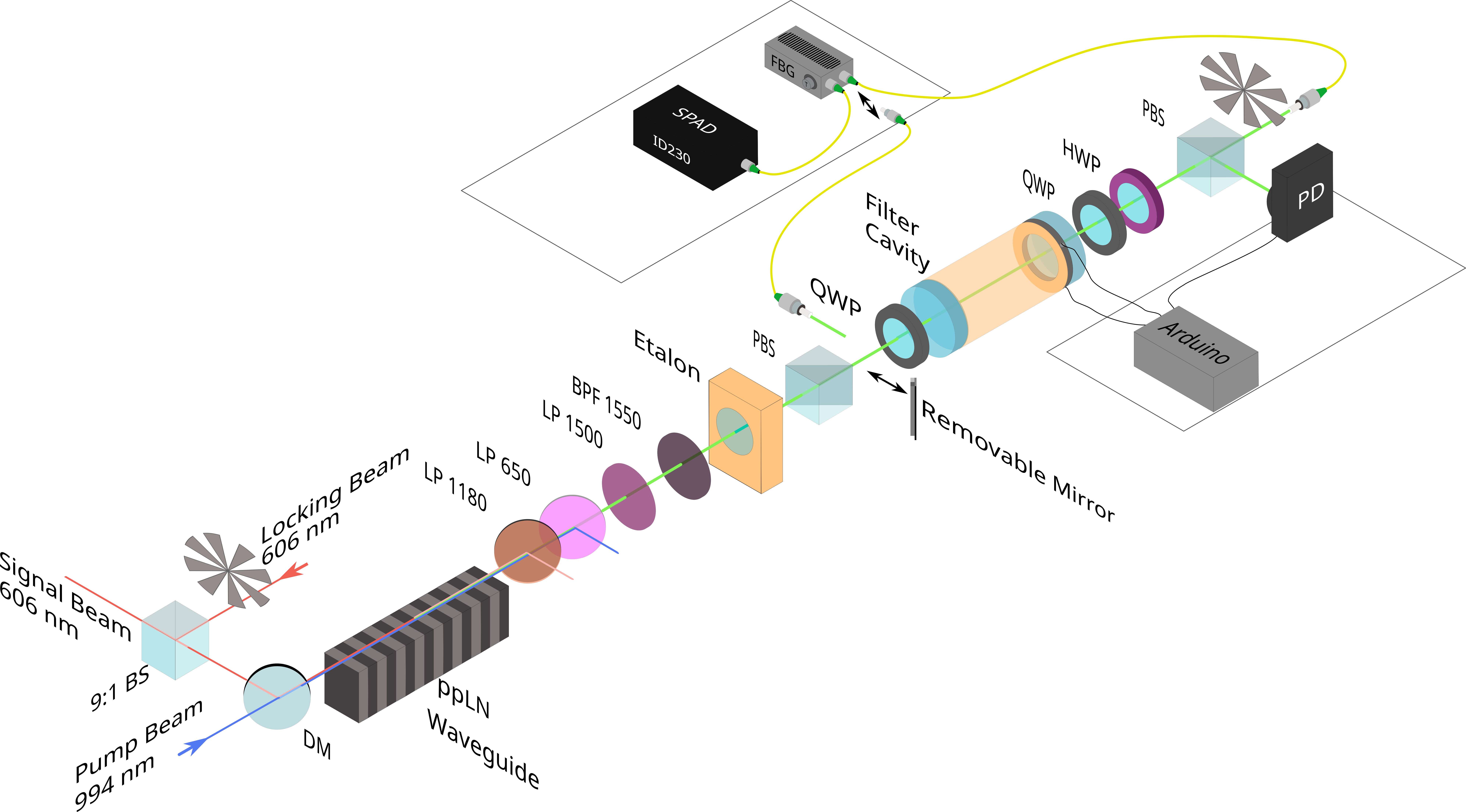}
\caption{\label{fig:setup}Schematic of the setup: In a periodically poled Lithium Niobate (MgO:ppLN) ridge waveguide we used difference-frequency conversion (DFG) to convert photons from \qty{606}{\nm} to \qty{1552}{\nm}. The pump beam and signal beam are overlapped by means of a dichroic mirror (DM) and coupled into the waveguide. After the waveguide, two color-glass filters (LP1180 and LP650) with a cut-on around \qty{1180}{\nm} and \qty{650}{\nm}  remove the pump light and depleted signal pulses. Further filtering is applied to remove the weakly phase-matched SPDC photons around the converted signal. For this, narrow filtering is performed by a temperature-stabilized etalon (FWHM  \qty{210}{\MHz}) and an actively stabilized Fabry-Perot cavity (FWHM \qty{12.5}{\MHz}, Finesse 100). A further long-pass filter with a cut-on of around \qty{1500}{\nano\meter} (LP1500) was used in combination with a band-pass filter around \qty{1550}{\nano\meter} (BPF1550, FWHM \qty{9}{\nano\meter}), followed by a PBS and a quarter-wave plate (QWP) in order to suppress reflections between the etalon and the Fabry-Perot cavity. Further, a temperature-stabilized Fiber-Bragg grating (FBG, FWHM \qty{2.4}{\GHz}) is used. The input beam is alternating due to the chopper wheel at  \qty{30}{\Hz} between  strong laser light of \qty{100}{\uW} in CW and a signal of weak attenuated pulses. The second chopper protects the single photon detector (SPAD) during the locking time and opens it during the measurement time. 
Via a removable mirror, the Fabry-Perot cavity could be bypassed in order to compare the performance of the setup with and without the narrow filtering. Further abbreviations: PD: Amplified InGaAs photo detector; HWP: half-wave plate; 9:1 BS: beam-splitter with splitting ratio 9:1.   }
\end{figure*}
The main challenges that hinder the use of QFC based on difference-frequency conversion as a tool to convert long single photons from the visible to the telecom C band are given by the noise produced in the conversion process. 
If the signal wavelength is smaller than \qty{775}{\nm},  the pump beam will have a shorter wavelength than the one of the converted signal, which is around \qty{1550}{\nm}. In this case, the noise is dominated by weakly phase-matched spontaneous parametric down-conversion (SPDC) from random domain errors in the waveguide poling~\cite{pelc2010influence}.

This noise can be reduced using different approaches. For broadband photons, temporal gating can be used. However, for narrowband photons that can interact with atomic systems, other techniques must be used, for example  narrow filtering~\cite{albrecht2014waveguide,farrera2016nonclassical,ikuta2018,krutyanskiy2017, Maring2017photonic,dreau_quantum_2018,bock2018,yu2020entanglement,vanleent2020longdistance,VanLeent2022,Knaut2024}, by carefully choosing a material (KTP) which reduces the produced noise~\cite{mann_low-noise_2023}, and by using a nonlinear crystal without poling~\cite{Geus2024}, reducing the produced noise even further. Another approach is to employ a two-step conversion process~\cite{pelc_cascaded_2012,esfandyarpour_cascaded_2018,Schaefer2024}. In this case the last conversion process happens by means of a pump beam at a longer wavelength than the target wavelength such that the noise photons are not in the same spectral region as the converted photons. 

In this work, we present the quantum frequency conversion of weak coherent pulses at the single photon level of up to \qty{13.9}{\us} duration full-width at half maximum (FWHM) from \qty{606}{\nm} to \qty{1552}{\nm} by means of a pump laser around \qty{994}{\nm}. Using ultra-narrowband spectral filtering down to \qty{12.5}{\MHz} with a Fabry-Perot cavity, we achieve a noise floor of \qty{300}{\per\second} after the waveguide, and a device efficiency of about \qty{25}{\percent}, leading to a signal-to-noise ratio $>$\num{460} for \qty{10}{\micro\second}-long pulses containing one photon on average.

\section{Experimental Setup and Procedure}
The experimental setup is depicted in Figure~\ref{fig:setup}. In a periodically poled MgO-doped Lithium-Niobate ridge waveguide (NTT), weak coherent pulses from a stabilized \qty{606}{\nano\meter} laser were overlapped with a strong \qty{994}{\nano\meter} pump laser with power $P_p$ in order to employ difference-frequency conversion to convert the pulses to \qty{1552}{\nano\meter}. Unless indicated otherwise, the pump power used for the various experiments was \qty{200}{\mW}, measured at the input of the waveguide. The coupling efficiency of the pump beam in the waveguide is 90 $\%$.  On a 9:1 beam-splitter, the signal beam was overlapped with a stronger beam, continuous but chopped at \qty{30}{\hertz}, used as locking beam. Each cycle (\qty{33.3}{\ms}) consists of   \qty{9}{\ms} locking time during which classical light is being sent and converted to lock the Fabry-Perot cavity. In the remaining part of the cycle, weak coherent pulses were sent. Before and after the locking time, a transition period of 50 pulses  was left to account for instabilities in the chopper synchronizations and the time it takes the chopper to block and unblock the beam. This leaves a measurement time of \qty{19.3}{\ms} in the case of short pulses and \qty{9.5}{\ms} in the case of long pulses. 

After conversion, the converted photons at telecom wavelength pass two long-pass colour-glass filters with cut-off wavelengths at  \qty{650}{\nano\meter} and \qty{1180}{\nano\meter}, respectively. This is followed by a temperature-stabilized etalon with a linewidth of \qty{210}{\mega\hertz} and a finesse of 19. To increase the signal-to-noise ratio, the photons can then be further filtered using an actively stabilized Fabry-Perot cavity with a linewidth of \qty{12.5}{\MHz} and a finesse of 100.  The cavity and etalon are both actively temperature-controlled. Finally, a temperature-stabilized Fiber-Bragg grating (FWHM \qty{2.4}{\GHz}) is used to further suppress the longitudinal modes of the cavity. 

\begin{figure}[htpb]
\includegraphics[width=1.0\columnwidth]{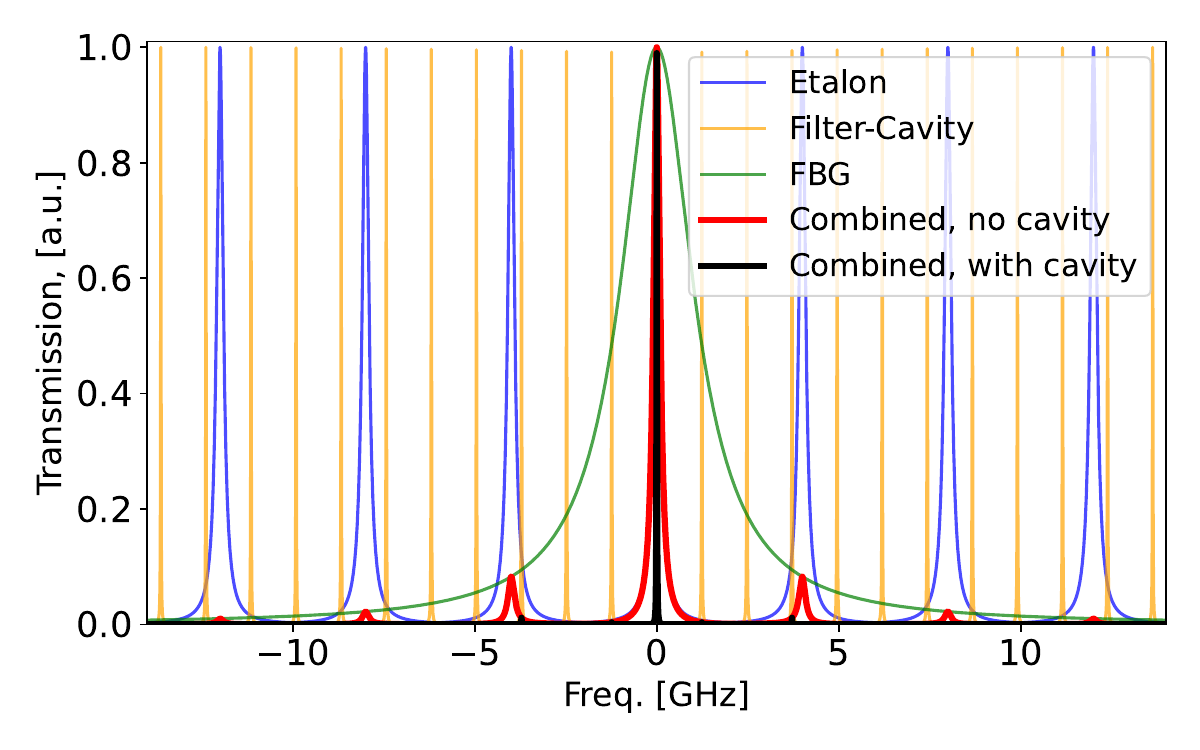}
\caption{\label{fig:filteringsystem}Simplified model  of the transmission spectrum of the etalon, filtering cavity and Fiber-Bragg grating. The transmission functions of the respective filtering elements have been multiplied to yield the purple and the brown trace. The ratio between the integrals over the combined transmission function with and without filter cavity is about 20.1.}
\end{figure}

During the measurement phase, the strong light for locking the cavity is blocked, and the weak attenuated pulses are converted to the telecom band and being detected by the SPAD (ID230). The signals from the SPAD were registered by a time-tagger (qutools QTAU) together with a TTL which accompanies every weak pulse. A schematic overview of the transmission functions through the filtering system and their respective combinations with and without the filtering cavity is depicted in figure~\ref{fig:filteringsystem}. The ratio between the areas under the combined transmission functions with and without filter cavity is about 20.1.
During the locking phase, the second chopper wheel blocks the converted light in front of the detector and the photodetector can read a classical light signal and transmit the output to the Arduino microcontroller which controls the position of the piezo of the cavity. The locking mechanism follows a simple maximization algorithm to keep the cavity locked on resonance. 
The same AOM was used to create the Gaussian-shaped pulses and the continuous light during the locking phase. This AOM was controlled by an AWG module (Signadyne) in a PXIe-chassis which, in turn, received a trigger signal from the Arduino microcontroller which was synchronized with the chopper controller.

\section{Results}

\begin{figure}
\includegraphics[width=1\columnwidth]{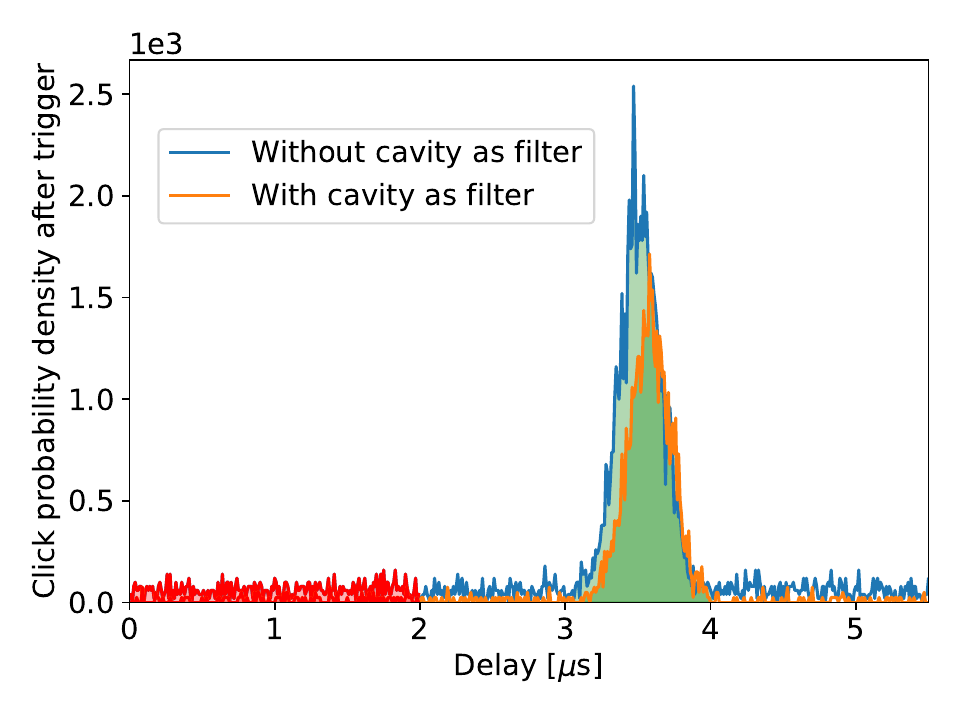}
\caption{\label{fig:hist_example_shortpulse} Detection probability density after the trigger corresponding to each weak coherent pulse. In this case, $\mu_{in}=0.021\pm0.002$ for a FWHM of $\qty{385}{\ns}$, and $P_p$=\qty{200}{\mW} before the waveguide. The Fabry-Perot filter cavity  increases the signal-to-noise ratio from $7.0\pm0.3$ to $98 \pm 22$. This means $\mu_1$ decreased from $\mu_1=\num[separate-uncertainty=true]{3.0(3)e-3}$ to $\mu_1=\num[separate-uncertainty=true]{2.1(0.5)e-4}$.
 The red shaded area is the area that has been considered to estimate the noise floor, the green area is the region of the pulse that was considered, corresponding to a 2.5-fold of the FWHM. For the  values stated above, dark counts were subtracted. Without this correction, the SNR increases from $6.1 \pm 0.3$ to $21 \pm 2$.}
\end{figure}

First, we convert short pulses at the single photon level. We record histograms  of the photon arrival times with respect to a trigger signal sent with each of the weak pulses. An example histogram for a short pulse of $\qty{385}{\nano\second}$ and  mean input photon number \mbox{$\mu_{in}=0.021\pm0.002$} can be seen in figure~\ref{fig:hist_example_shortpulse}. From this kind of histogram, we infer the signal-to-noise ratio of the converted photon by comparing the recorded counts in the pulse with a noise window on the side. The size of the detection window is 2.5 times the FWHM of the input pulse. As figure of merit, we use the parameter $\mu_1=\mu_{in}/SNR$, which represents the number of input photons to reach a SNR=1 after the conversion. Conversely, $1/\mu_1$ gives the SNR for an input pulse containing 1 photon on average. For that measurement, without the filtering optical cavity the SNR is $7.0 \pm 0.3$ , leading to $\mu_1=\num[separate-uncertainty=true]{3.0(3)e-3}$. Including the optical cavity increases the SNR to to $98 \pm 22$ ($\mu_1=\num[separate-uncertainty=true]{2.1(0.5)e-4}$).

\begin{figure}
\includegraphics[width=1\columnwidth]{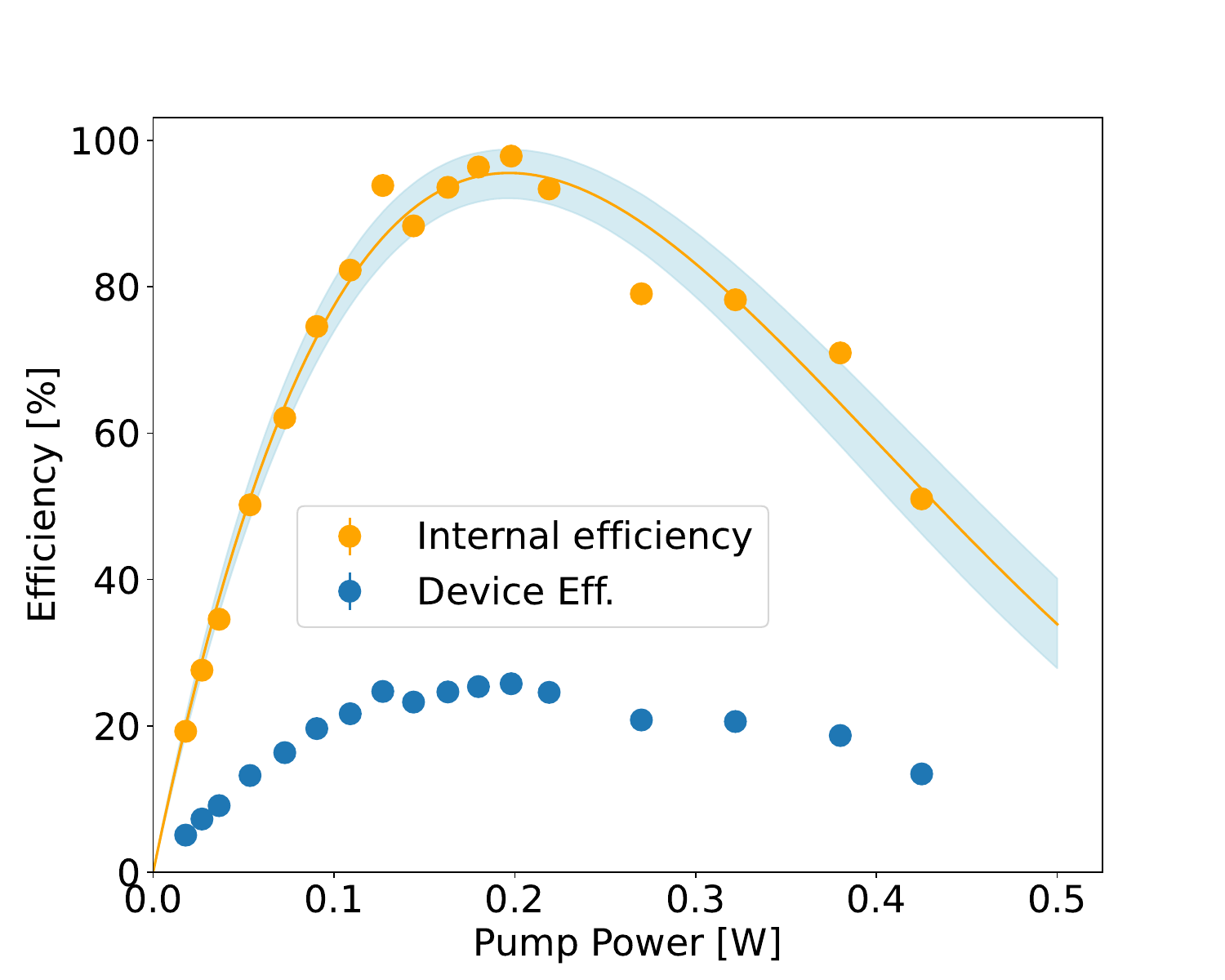}
\caption{\label{fig:eff_vs_power}Device efficiency (blue) as a function of the pump power coupled into the waveguide. The device efficiency corresponds to the probability that a photon in front of the waveguide is converted and arrives at the detector. The internal efficiency (orange) was calculated correcting for the losses due to waveguide and fiber couplings and filtering transmission. The FWHM pulse length was \qty{1170}{\nano\second} with a mean photon number per pulse of \num[uncertainty-mode=separate]{0.07(1)}. The shaded region corresponds to a confidence interval of one standard deviation estimated with a Monte Carlo simulation.}
\end{figure}

We then characterize the system efficiency. Fig. ~\ref{fig:eff_vs_power}, shows the device and internal efficiency measured with the filtering cavity as a function of the pump power $P$, using input pulses with a FWHM length of \qty{1.17}{\micro\second}. It is fitted to a function following the model proposed in~\cite{Albota:04}.
\begin{equation}
    \eta = \eta_{\mathbf{max}}\sin^2(L\sqrt{\beta P}).\label{eq:eff}
\end{equation}

$L$ corresponds to the length of the waveguide, \qty{48}{\mm} in our case.
The fit yields a normalized efficiency $\beta$ of \mbox{\qty[uncertainty-mode=separate]{54\pm3} {\percent\per\W\per\cm\squared} }and a maximum internal efficiency of \mbox{\qty[uncertainty-mode = separate]{95\pm3}{\percent}}. 

The highest device conversion efficiency was \qty[uncertainty-mode=separate]{25.8(0.3)}{\percent} at \qty{198}{\mW} pump power. The device efficiency represents the probability that a 606-nm-photon in front of the waveguide was converted and collected into the fiber that goes to the SPAD. It has been calculated by dividing the click probability per trigger by the mean input photon number of \num[uncertainty-mode=separate]{0.07(1)} per pulse given from a power-meter measurement and calibrated neutral density filters (\qty{-93}{\dB}), and then corrected for the \qty{10}{\percent} detection efficiency of the SPAD.  The internal conversion efficiency is estimated by correcting also for the other losses that the converted signal is subjected to: the fiber coupling (about \qty{80}{\percent}) and the transmission in the filtering system (about \qty{48}{\percent}). This yields an efficiency of \qty{38}{\percent} for a converted photon after the waveguide to reach the detector. We further corrected for the coupling of the \qty{606}{\nm} photons into the waveguide (\qty{70}{\percent}) and assumed that losses inside the waveguide are negligible.  
While both the Fabry-Perot cavity as well as the etalon have a transmission of around \qty{90}{\percent}, the Fiber-Bragg grating has an efficiency of only \qty{60}{\percent}.

\begin{figure}
\includegraphics[width=1\columnwidth]{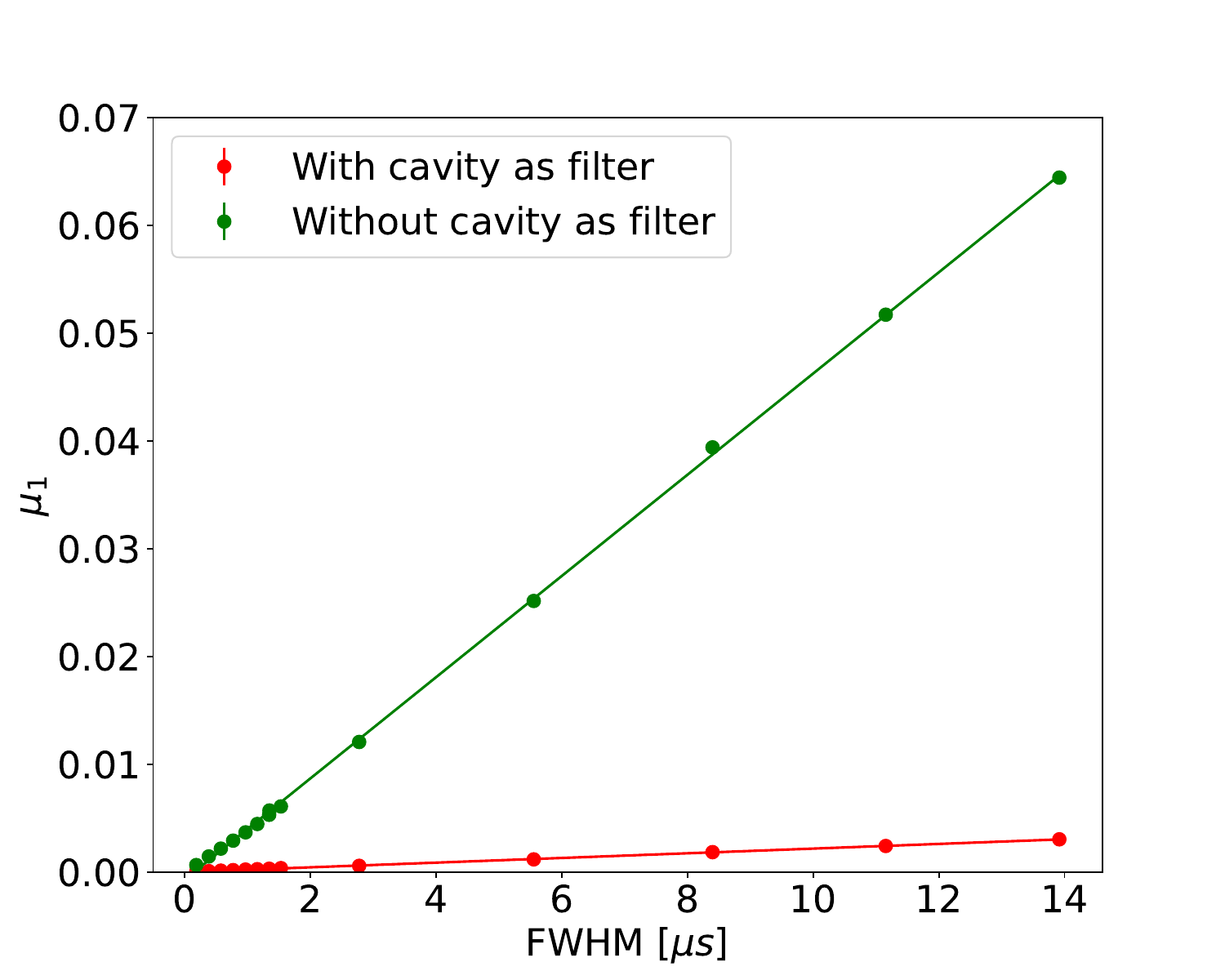}
\caption{\label{fig:mu1_vs_fwhm}Measurement of $\mu_1$ for different input pulse lengths, with $P_p$ =\qty{200}{\mW} before the waveguide. 
A linear fit shows that $\mu_1$ grows with slope of 
\num[separate-uncertainty = true]{2.17(1)e-4}~\si{\per\micro\second} in the case with the filtering cavity and 
\num[separate-uncertainty = true]{4.70(2)e-3}~\si{\per\micro\second} in the case without the filtering cavity. 
The size of the detection window is 2.5 times the FWHM of the input pulse length.  For the first 8 measurements with relatively short pulses, the pulses were sent at a repetition rate of \qty{20}{\kilo\hertz} and in the case of the longest pulses, this resulted in a detection probability of 
\num{2.7e-3}. Due to the duty cycle of the chopper, 11600 pulses per second were used for the data analysis. In total,  between \num{2e6} and \num{5e6} pulses were used for each measurement. For the longer pulses after the first eight measurements, the spacing of the pulses was increased and the pulse rate was therefore only about \qty{7.1}{\kilo\hertz}, out of which 2000 pulses per second were used for the analysis. The data presented consists of two measurement series. The first 8 measurements for relatively short pulses and the next 10 measurements for which the measurement sequence and rate had to be changed due to the long pulses. For each of the two measurement series, the first measurement (shortest pulse) was used to define the noise-level for the whole series.  }
\end{figure}

\begin{figure}
\includegraphics[width=1\columnwidth]{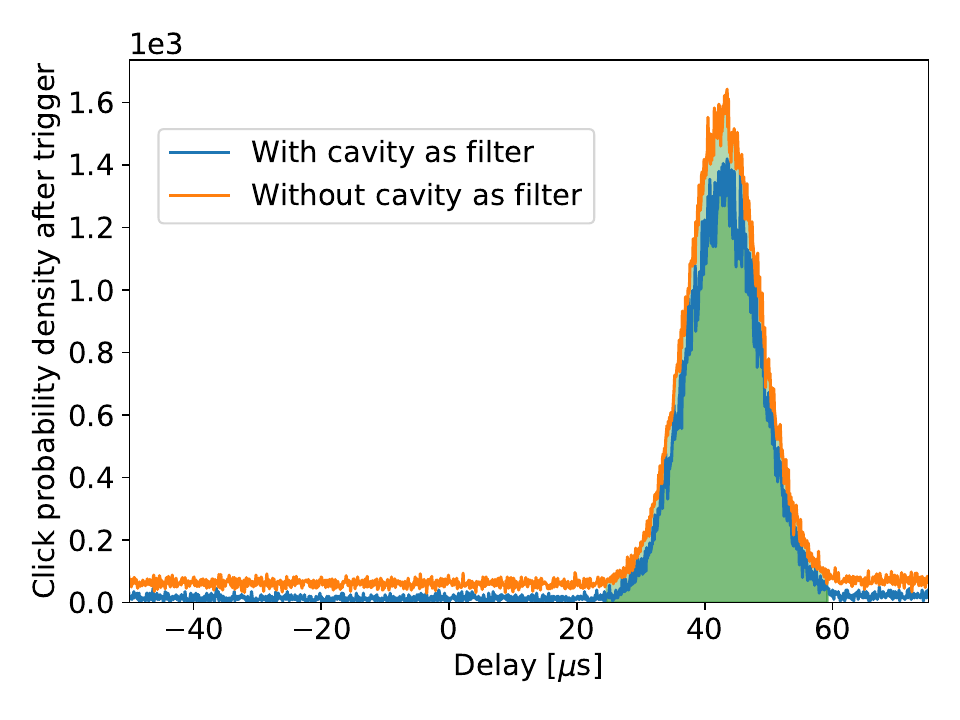}
\caption{\label{fig:hist_example} Detection probability density after the trigger corresponding to each weak coherent pulse. In this case, $\mu_{in}=0.77\pm0.08$ for a FWHM of $\qty{13.9}{\us}$, and $P_p$ = \qty{200}{\mW} before the waveguide. The Fabry-Perot cavity filter increases the signal-to-noise ratio from $13.7\pm0.2$ to $293\pm 26$. This decreased $\mu_1$ from $\mu_1=\num[separate-uncertainty=true]{6.0(6)e-2}$ to $\mu_1=\num[separate-uncertainty=true]{2.6(3)e-3}$. 
 The green shaded area is the region of the pulse that was considered, corresponding to a 2.5-fold of the FWHM. For the  values stated above, dark counts were subtracted. Without this correction, the SNR increases from $11.7\pm0.1$ to $50\pm2$.}
\end{figure}

Finally, we investigated how the conversion process behaves as a function of pulse length. 
The measurement reported in figure~\ref{fig:mu1_vs_fwhm} shows how much $\mu_1$ increases with increasing pulse length. The noise level increases with pulse duration because  the detection window has to be increased. However, the efficiency of the conversion process was  not significantly affected, since the waveguide allows for conversion within \qty{24}{\THz} FWHM within the current setup while the shortest pulses we sent have a spectral width of \qty{2.4}{\MHz}.  
Because more noise is collected, the signal-to-noise ratio decreases with longer pulses. The plot contains a comparison of the case with the Fabry-Perot filtering cavity and without. The mean photon number for these pulses changed from 0.01 photons per pulse for the shortest pulse (FWHM \qty{186}{\nano\second}) to 0.8 photons per pulse for the longest pulses (\qty{13.45}{\micro\second}). The peak power of the signal pulses was not adjusted during the measurement, but taken into account in post-processing. The highest observed click rate was about \qty{210}{\per\second}, therefore we are confident that the InGaAs SPAD with a dead-time set to \qty{20}{\micro\second} and an efficiency of \qty{10}{\percent} was not used in a regime where it shows saturation effects. Since the observed noise was dominated by dark counts, which could be avoided by using a superconducting nano-wire single photon detector, we subtracted a dark count rate of \qty{9.13}{\per\second} in the analysis.

A linear fit shows that $\mu_1$ grows with the proportionality constant of 
\num[separate-uncertainty = true]{2.17(1)e-4}~\si{\per\micro\second} in the case with the filtering cavity and 
\num[separate-uncertainty = true]{4.70(2)e-3}~\si{\per\micro\second} in the case without the filtering cavity.  Without subtracting dark counts, the values of the slopes are 
\num[separate-uncertainty = true]{1.151(4)e-3}~\si{\per\micro\second} and 
\num[separate-uncertainty = true]{5.47(2)e-3}~\si{\per\micro\second} for the case with and without the cavity, respectively. The dark-count subtracted values show a noise reduction of a factor of 21.6 when inserting the filter cavity, consistent with a simple numerical model of the filtering system assuming a flat noise spectrum which predicts a noise reduction by a factor 20.1 by combining the etalon (FWHM \qty{210}{\MHz}) with the filter cavity (FWHM \qty{12.5}{\MHz}) as displayed in figure~\ref{fig:filteringsystem}. 
Another example histogram for the longest pulses converted (FWHM of \qty{13.9}{\micro\second}) can be seen in figure~\ref{fig:hist_example}.

\section{Discussion and Outlook}
We now analyze the results in the context of the conversion of non-classical states emitted by a quantum memory, and in particular to which extent the demonstrated SNRs would preserve quantum correlations in a realistic quantum memory setup. It has been shown that if one photon from a photon pair is frequency converted, the second order cross-correlation between signal and idler $g^{(2)}_{s,i}$ evolves as~\cite{albrecht2014waveguide}: 
\begin{equation}
    g^{(2)}_{c,i}=g^{(2)}_{s,i}\frac{\mu_{in}/\mu_1+1}{\mu_{in}/\mu_1+g^{(2)}_{s,i}}
\end{equation}

This model predicts that given the parameters of the experiment reported in ref.~\cite{kutluer2017solidstate},  the measured cross-correlation of $g^{(2)}_{s,i}= 17.3$ at a detection window size of \qty{600}{\nano\second} and an efficiency of \qty{1.6}{\percent} would only reduce to 15.3 with this conversion setup. 

As mentioned in the introduction, an important use case of our converter would be to convert long photons emitted from the Praseodymium memory such that they can interfere with photons emitted by trapped ions. If we consider photon durations of \qty{10}{\micro\second}, corresponding to a linewidth of \qty{44}{\kilo\hertz}  for Gaussian pulses and a retrieval efficiency of \qty{10}{\percent} from the memory, a maximum $g^{(2)}_{c,i}$ of 47 can be reached. 

A possible improvement of the setup would be to replace the lossy Fiber-Bragg grating with a volume-holographic Bragg grating in combination with a pi-shifted Fiber-Bragg grating in order to select one or a few spectral modes of the Fabry-Perot cavity. Another possible way would be to use a very thin etalon with a very large free spectral range, such that no additional modes are within bandwidth of the Bragg grating~\cite{hellerlukas23}.

We presented a setup which can be used to convert long photons from, for example, praeseodymium-based DLCZ-AFC experiments or other solid-state quantum memories to the telecom band. This constitutes an important step towards realizing quantum repeaters with praeseodymium-based quantum memories. Also, it shows the possibility of interfacing solid-state quantum repeater nodes with processing nodes like cold calcium ions which emit photons of durations similar to the longest photons we converted here~\cite{krutyanskiy2023}. 

\section*{Acknowlegement}
We thank Alessandro Seri, Bernardo Casabone and Josep-Maria Batllori Berenguer for their help in the early stages of the project. This project received funding from the European Union research and innovation program within the Flagship on Quantum Technologies through Horizon 2020 grant 820445 (QIA) and  Horizon Europe project QIA-Phase 1 under grant agreement no. 101102140, from the Government of Spain (PID2019-106850RB-I00; Severo Ochoa CEX2019-000910-S; BES-2017-082464), from the MCIN with funding from European Union NextGenerationEU PRTR (PRTR-C17.I1), from the Gordon and Betty Moore Foundation through Grant No. GBMF7446 to H. d. R, from Fundaci\'o Cellex, Fundaci\'o Mir-Puig, and from Generalitat de Catalunya (CERCA, AGAUR).

\providecommand{\newblock}{}

\providecommand{\newblock}{}

\end{document}